\documentclass[useAMS,usenatbib]{mn2e}
\usepackage{graphicx,amssym}
\newcommand{\bc}{\begin{center}}
\newcommand{\ec}{\end{center}}

\title[Galaxy formation in the 1st and 7th-year WMAP cosmologies]
      {Galaxy formation in WMAP1 and WMAP7 cosmologies}
\author[Qi Guo et al.]
       {\parbox{18cm}{Qi Guo$^{1,2,3}$
           \thanks{Email:guoqi@nao.cas.cn}, Simon White$^{2}$, Raul
           E. Angulo$^{2}$,  Bruno Henriques$^{2}$, Gerard
           Lemson$^{2}$, Michael Boylan-Kolchin$^4$, Peter Thomas$^5$,
         Chris Short$^5$}
       \\     
       \\
       $^{1}$Partner Group of the Max-Planck-Institut f\"ur
       Astrophysik, National Astronomical Observatories, Chinese
       Academy of Sciences, \\Beijing, 100012, China \\
       $^{2}$ Max Planck Institut f\"ur					       
         Astrophysik, Karl-Schwarzschild-Str. 1, 85741 Garching, Germany\\  
	$^3$ Institute for Computational Cosmology, Department of Physics, University of Durham, South Road, Durham, DH1 3LE, UK \\ 
        $^4$ Center for Cosmology, Department of Physics and
        Astronomy, 4129 Reines Hall, University of California, Irvine,
        CA 92697, USA  \\
        $^5$ Astronomy Centre, University of Sussex, Falmer, Brighton BN1 9QH
      }
\begin{document}

\date{Accepted 2012 October 1. Received 2012 September 25; in original form 2012 May 31}

\volume{428}
\pagerange{1351--1365} 
\pubyear{2013}

\maketitle

\label{firstpage}

\begin{abstract}

Using the technique of Angulo \& White (2010) we scale the Millennium
and Millennium-II simulations of structure growth in a $\Lambda$CDM
universe from the cosmological parameters with which they were carried
out (based on first-year results from the Wilkinson Microwave
Anisotropy Probe, WMAP1) to parameters consistent with the seven-year
WMAP data (WMAP7).  We implement semi-analytic galaxy formation
modelling on both simulations in both cosmologies to investigate how
the formation, evolution and clustering of galaxies are predicted to
vary with cosmological parameters.  The increased matter density
$\Omega_m$ and decreased linear fluctuation amplitude $\sigma_8$ in
WMAP7 have compensating effects, so that the abundance and clustering
of dark halos are predicted to be very similar to those in WMAP1 for
$z\leq 3$. As a result, local galaxy properties can be reproduced
equally well in the two cosmologies by slightly altering galaxy
formation parameters. The evolution of the galaxy populations is then
also similar. In WMAP7, structure forms slightly later. This shifts
the peak in cosmic star formation rate to lower redshift, resulting in
slightly bluer galaxies at $z=0$.  Nevertheless, the model still
predicts more passive low-mass galaxies than are observed.  For $r_p<
1$~Mpc, the $z=0$ clustering of low-mass galaxies is weaker for WMAP7
than for WMAP1 and closer to that observed, but the two cosmologies
give very similar results for more massive galaxies and on large
scales. At $z>1$ galaxies are predicted to be more strongly clustered
for WMAP7. Differences in galaxy properties, including, clustering, in
these two cosmologies are rather small out to $z\sim 3$. Given that
there are still considerable residual uncertainties in galaxy formation
models, it is very difficult to distinguish WMAP1 from WMAP7 {through observations of galaxy properties or their evolution.} 
\end{abstract}


\section{Introduction}

In the standard picture, galaxies form through the condensation of gas
at the centres of a hierarchically aggregating population of dark
matter haloes.  The pattern of halo evolution is controlled by the
statistics of the primordial density fluctuations which emerged from
the early universe, and by the global cosmological parameters which
determine their growth rate at late times. Galaxy evolution depends in
addition on the interplay between gas inflows, star formation,
radiative, chemical and hydrodynamical feedback, and dynamical
processes such as merging and tidal disruption.  The intrinsic
properties and the spatial distribution of galaxies are thus closely
related both to cosmological issues such as the composition and early
evolution of the universe, and to astrophysical issues such as the
generation and consequences of galactic winds. As a result, the
precision with which cosmological conclusions can be drawn from the
new generation of very large galaxy surveys will depend on the extent
to which the relevant signals are distorted by the astrophysics of
galaxy formation.

A straightforward approach to exploring how cosmological inferences
are limited by our understanding of galaxy formation is to take
large-volume, high-resolution simulations of cosmic structure
formation in various cosmologies, to populate each with galaxies using
a broad variety of physically and observationally consistent
treatments of the baryonic processes, and to use mock catalogues to
study whether real surveys can distinguish the effects of cosmological
and galaxy formation parameters.  Direct hydrodynamical simulations do
not yet produce galaxy populations with basic statistics (e.g. stellar
mass functions) consistent with observation, and are in any case too
expensive for even a single simulation to be carried out over a volume
approaching those of current surveys. Halo Occupation Distribution
(HOD) models can be applied to very large simulations, producing
galaxy populations with abundance and correlation statistics in close
agreement with observation, but it is unclear whether their
simple, purely statistical assumptions lead to physically consistent
predictions at different redshifts, or represent adequately the relevant aspects
of galaxy formation. Semi-analytic simulations are equally easily
constructed and fit galaxy abundances and clustering almost as
well as HOD models. Their schematic but plausible representation of formation
processes guarantees physically consistent populations at different
redshifts, and thus enables direct tests of the influence of
individual processes on the large-scale galaxy distribution.

N-body simulations large enough to represent current and
next-generation surveys at the resolution required for semi-analytic
galaxy formation modelling are still computationally
expensive \citep{Angulo2010}. As a result, it is not yet feasible to
carry out a large suite of simulations scanning the allowed
cosmological parameter space. In the current paper, we use the
rescaling technique of \cite{Angulo2010} to map the Millennium
Simulation \citep[MS:][]{Springel2005} and the Millennium-II
Simulation \citep[MS-II:][]{Boylan2009} from their original WMAP1
cosmology to one with the parameters preferred by the 7-year WMAP
results \citep{Komatsu2011}. Our aim is to test whether such scaling,
applied to a single simulation, can represent cosmic structure
sufficiently accurately to build reliable galaxy catalogues as
cosmological parameters are varied throughout the allowed range. We
show that the relevant statistical properties of (sub)haloes in the
scaled model are indeed very close to those in a simulation carried
out directly in the WMAP7 cosmology. Then, we also use such scaling to
explore how cosmology affects our predictions for the formation of
galaxies.

Our semi-analytic modelling is almost identical to that
in \cite{Guo2011}\footnote{A few minor bugs have been found and
corrected since this paper was published, but none of them changes its
results significantly.}.  The SAM follows gas infall (both cold and
hot, primordial and recycled), shock heating, cooling, star formation,
stellar evolution, supernova feedback, black hole growth, AGN
feedback, metal enrichment, mergers, and tidal and ram-pressure
stripping. Galaxy formation and evolution are followed from $z>10$ to
the present. As in \cite{Wang2008}, who used smaller simulations and
an older galaxy formation model, we adjust model parameters
independently in each cosmology to reproduce observations of the
low-redshift galaxy population, in our case, SDSS stellar mass and
luminosity functions, colour distributions, metallicities and gas
fractions.  We then study differences in the implied clustering and
evolution to see if these are clearly related to the difference in
cosmological parameters.

\cite{Wang2008} compared results in the WMAP1 and WMAP3
cosmologies. Structure formation differs significantly more between
these than between WMAP1 and WMAP7. The most relevant parameters are
$\sigma_8$ and $\Omega_m$. WMAP3 advocated a much lower $\sigma_8 =
0.7$ and also a lower $\Omega_m = 0.23$ than WMAP1 (0.9 and 0.25,
respectively) while WMAP7 prefers an intermediate $\sigma_8 = 0.8$,
and a higher $\Omega_m = 0.27$. As a result, structure formation in
the WMAP7 cosmology is considerably closer to that in WMAP1 than to
that in WMAP3.  We will see that, at the current level of precision,
predictions for the galaxy populations in WMAP1 and WMAP7 are
difficult to distinguish.

This paper is organised as follows. In Sec.\ref{sec:sim} we briefly
describe the MS and MS-II which provide our halo/subhalo merger trees.
We compare (sub)halo abundances and clustering as functions of
redshift in the two cosmologies, as well as comparing results from the
scaled MS to results from a similar simulation carried out directly in
the WMAP7 cosmology. This section also describes particularly relevant
physical recipes from the galaxy formation model.  In
Sec.\ref{sec:result} we present abundances, scaling relations and
clustering properties for low-redshift galaxies in the scaled (WMAP7)
simulations and compare them to those in the original unscaled (WMAP1)
case. This section also presents a comparison of the evolution of the
galaxy populations in the two cosmologies. We summarise our main
results and discuss the future application of these techniques in
Sec.\ref{sec:conclusion}. The WMAP7 galaxy catalogues associated with
this paper are made publicly available with its acceptance on the same
site\footnote{http://www.mpa-garching.mpg.de/millennium} and in the
same format as the previously released catalogues for the WMAP1
cosmology.

\section{  N-body simulations and semi-analytic models}
\label{sec:sim}

\subsection{N-body simulations} 

We simulate the evolution of the galaxy population by implementing the
galaxy formation model of \cite{Guo2011} on subhalo merger trees
extracted from three large cosmological N-body simulations; the
Millennium Simulation \citep[MS:][]{Springel2005}, the Millennium-II
Simulation\citep[MS-II:][]{Boylan2009} and a simulation identical to
the MS {(including the same box-size, $500h^{-1}$Mpc, output
redshift sequence and post-processing pipeline)} but with cosmological
parameters consistent with the latest observational constraints
(MS-W7: Thomas et al., in preparation).  In Table 1 we provide the
numerical values of the most relevant parameters. All three
simulations follow 2160$^3$ particles from redshift 127 to the present
day. The MS and the MS-W7 were carried out in a box of side 500 Mpc/h,
whereas the MS-II used a box of side 100 Mpc/h. The mass of the
simulation particles in these runs is $m_p = 8.61\times10^8$ (MS),
$6.88\times10^6$ (MS-II) and $9.31\times10^8 M_{\odot}$ (MS-W7). {The MS-W7 was carried out with different fluctuation phases from the
MS so that cosmic variance affects any comparison of rare objects
between the two simulations.}

In addition, we will employ two extra catalogues, with identical
cosmological parameters to those of the MS-W7, generated from
the MS and MS-II using the scaling algorithm developed by 
\cite{Angulo2010}. We will refer to them as MS-SW7 and MSII-SW7,
respectively. This algorithm allows scaling of the results of an
N-body simulation from its original cosmology to another with modified
parameters, and it involves three steps: rescaling the box length,
mass and velocity units, relabelling the output times, and rescaling
the amplitudes of individual large-scale (linear) fluctuation
modes. See Tabel 1 for the scalings required, and a few examples of
the relabelling of outputs. {Here we choose z = 0 in WMAP7 to
correspond to snapshot 53 ($z=0.28$) in the MS.} The scheme reproduces
dark halo masses, positions, velocities, and clustering in the target
cosmology to the few percent level at all
times \citep{Angulo2010,Ruiz2011}.  {Note that, because of the
relabelling, output redshifts other than $z=0$ do not correspond
exactly between the scaled and unscaled simulations. Note also that we
do not apply the final step of adjusting the amplitudes of large-scale
linear modes, because this has negligible effects on the clustering
statistics analysed in this paper.}

The most significant difference between the cosmologies preferred by
WMAP7 and WMAP1 data, is a 10\% lower value of $\sigma_8$. This
implies a lower amplitude for primordial density fluctuations, which
translates into a decrease in the number of haloes with masses above
$M_*$, and an increase for those below this characteristic mass. The
impact of $\sigma_8$ was partially compensated by a higher value of
$\Omega_m$ in WMAP7 compared to WMAP1. As a result, halo mass
functions are predicted to be very similar in the two cosmologies at
$z=0$ over the range of halo masses most relevant for galaxy
formation. In the next subsection we will explore in more detail how
these changes in cosmology affect aspects of dark matter structure
relevant for galaxy formation.

\begin{table}
\caption{Summary of the parameters adopted in our WMAP1 and WMAP7 cosmologies, as
  well as of the scale factors, $f_l$, $f_{mp}$ and $f_{m_{vir}}$ for
  simulation box length, particle mass and halo mass, respectively.
  Here $F$ is a function of (original) halo concentration parameter
  $c$ and of the ratio between the (dimensionless) matter density in
  the WMAP1 cosmology at the output redshift under consideration,
  $\Omega_o(z_o)$, and that for WMAP7 at the redshift to which this
  ouput maps, $\Omega_t(z_t)$. In addition we give the redshifts $z_o$
  in WMAP1 which are mapped to redshifts $z_t = 0$, $\sim 1$ and $\sim
  3$ in WMAP7}

\begin{tabular}{||l||l||c||} 

\hline
Parameter   & WMAP1 & WMAP7  \\
\hline
 $\Omega_{m}$       & 0.25 & 0.272\\
 $\Omega_{\Lambda}$     & 0.75 & 0.728 \\
  $\Omega_{b}$     & 0.0425 & 0.045  \\
 $h $   & 0.73 & 0.704\\
 $n$  & 1 & 0.961 \\
  $\sigma_{8}$    & 0.9 & 0.807\\
  $f_l$   & 1 & 1.043 \\
  $f_{mp}$  & 1 & 1.23488 \\ 
 $f_{m_{vir}}$  & 1 & $f_{mp} \times F(c,\Omega_o(z_o)/\Omega_t(z_t)$)\\
 redshift & 0.28 & 0\\
 &1.39&1.02\\
 &3.58&2.92\\

\hline
\end{tabular} 
\label{table:cospara}
\end{table}

\subsection{Dark matter halo and subhalo properties} 

 Galaxies form through the condensation of gas and the merging of
satellites, both of which are accreted along with the dark matter as
their surrounding haloes build up with time.  Their stellar mass is
thus closely related to the total mass of their haloes. The outer
parts of haloes are removed by tidal effects if they fall into larger
systems and so become satellite subhaloes, while the stars and gas of
their central galaxies are less easily stripped. Thus the
stellar mass of satellite galaxies is more closely linked to halo mass
at infall than it is to current subhalo mass \citep[see, for example,
the galaxy formation simulations
of][]{Springel2001,Gao2004,Guo2011}.
This realisation led to a simplified model which assumes a monotonic
relation between the stellar mass or luminosity of a galaxy and the
maximum mass or circular velocity ever attained by its
halo \citep[e.g.][]{Vale2004,Conroy2006,Moster2010,Guo2010}. This
scheme has proved successful in matching many local galaxy properties,
including the Tully-Fisher relation, galaxy correlation functions and
the observed stellar mass-halo mass relation. Its assumptions are at
least approximately obeyed by more physical models for the formation
and evolution of the galaxy
population \citep[e.g.][]{Guo2011} 

\begin{figure}
\bc
\hspace{-0.6cm}
\resizebox{8.5cm}{!}{\includegraphics{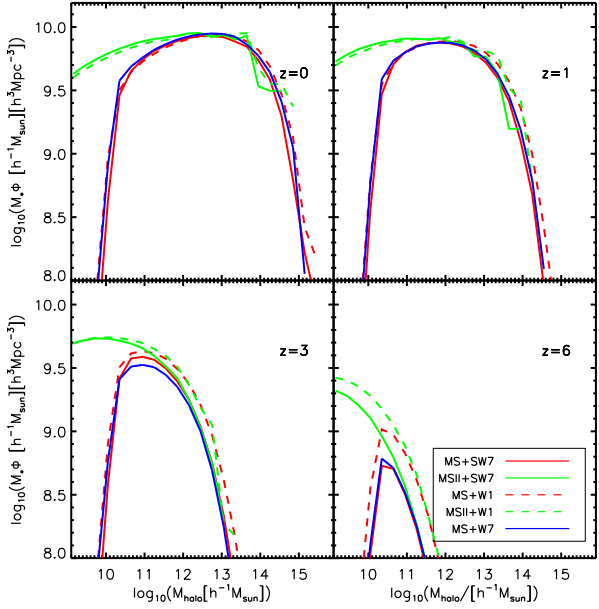}}\\%
\caption{Differential (sub)halo mass functions at $z\approx 0, 1, 3$ 
  and 6, 
  as indicated by the label in each panel. For the main subhalo of
  each FoF group this is the current $M_{200}$ of the group, while for
  satellite subhaloes it is the $M_{200}$ value just prior to first
  infall. Solid and dashed curves show measurements for the WMAP7 and
  the WMAP1 cosmology, respectively. Red, green and blue curves are
  results from the MS, the MS-II and the MS-W7, respectively.}
\label{fig:dmmass}
\ec
\end{figure}

Fig.~\ref{fig:dmmass} shows the abundance of dark matter (sub)haloes
 as a function of $M_{200}$, the mass within a sphere centred on the
gravitational potential minimum enclosing mean density 200 times the
critical value.  For central objects, this corresponds to its current
value, but for satellites, it corresponds to the value just prior to
(last) infall  onto the current host. We display results at redshifts 0,
1, 3 and 6. Solid and dashed curves refer to the WMAP7 and WMAP1
cosmologies, respectively, whereas coloured lines indicate our 
different simulations.

Results from the MS and MS-II converge well over the mass range
where both simulations have adequate resolution and good
statistics. This range shifts to lower mass at higher redshift as
satellites become progressively less important. At $z=0$ the
two simulations agree closely for masses above
$10^{12}M_{\odot}/h$. This is more than an order of magnitude larger
than the mass needed to get similar agreement for the standard halo
mass function \cite[e.g.][]{Boylan2009}. This reflects the inclusion
of subhalos; for the majority of subhalos to be reliably identified after infall and stripping they need to have $\sim 10^3$ particles
at infall.

 At $z=6$, (sub)haloes of all masses are more abundant at every mass in WMAP1 than in
WMAP7, but by $z=1$ the two mass functions are very close for
$M_{sh}<10^{12}M_{\odot}/h$ and at $z=0$, they overlap for
$M_{sh}<10^{12.5}M_{\odot}/h$, remaining close at higher masses although
there are always more high-mass haloes in WMAP1 than in WMAP7. The relatively small differences
between the two cosmologies, especially at low redshift, reflect the fact
that the lower value of $\sigma_8$ in WMAP7 is largely compensated by its
higher value of $\Omega_m$.

A comparison of the blue and red solid curves in Fig.~\ref{fig:dmmass}
shows that the (sub)halo mass functions for the scaled MS
 agree very well with those for the MS-W7 (which
was carried out directly in the WMAP7 cosmology) over the range not affected
by resolution. This confirms the results of \cite{Angulo2010} and
\cite{Ruiz2011}, and extends them to larger simulations and to
the (sub)halo mass function, which is the most relevant mass function for modelling of
the galaxy population. The differences seen at low masses, most
noticeably at  $z=3$, are partly due to fact that output times do not
coincide exactly between the two simulations, and partly to different mass resolutions, numerical settings and
initial phases. In addition, at these low masses many satellite haloes are
artificially lost in a way
that depends sensitively on the mass and force resolution of each simulation.

\begin{figure}
\bc
\hspace{-0.6cm}
\resizebox{8.5cm}{!}{\includegraphics{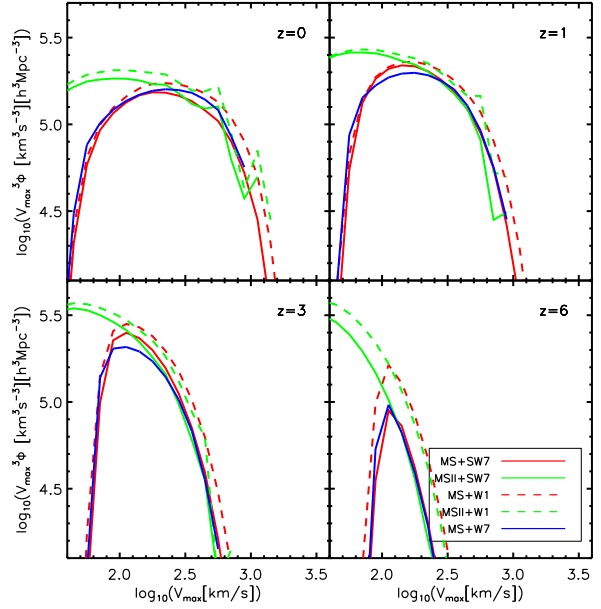}}\\%
\caption{Differential distributions of dark halo $V_{max,infall}$ as a function 
  of redshift. For the main subhalo of each FoF group this is taken to
  be its current maximum circular velocity, while for satellite
  subhaloes it is the value at infall.  the different panels, line
  colours and line types correspond exactly to those in
  Fig.~\ref{fig:dmmass}}
\label{fig:dmvmax}
\ec
\end{figure}

Haloes of given mass have higher maximum circular velocity, and hence
higher density and virial temperature, at higher redshift. This is
important for galaxy formation modelling because of its effect on gas
cooling rates.  Fig.~\ref{fig:dmvmax} is directly analogous to
Fig.~\ref{fig:dmmass}, differing only in that (sub)haloes are
characterised by their maximum circular
velocity rather than by their mass. Again the velocity is taken
to be the current value for main subhaloes and the value at infall for
satellite subhaloes. The behaviour here is very similar to that in
the earlier plot, with good convergence between MS and MS-II and
between the MS scaled to the WMAP7 cosmology and the MS-W7.
 
\begin{figure}
\bc
\hspace{-0.6cm}
\resizebox{8.5cm}{!}{\includegraphics{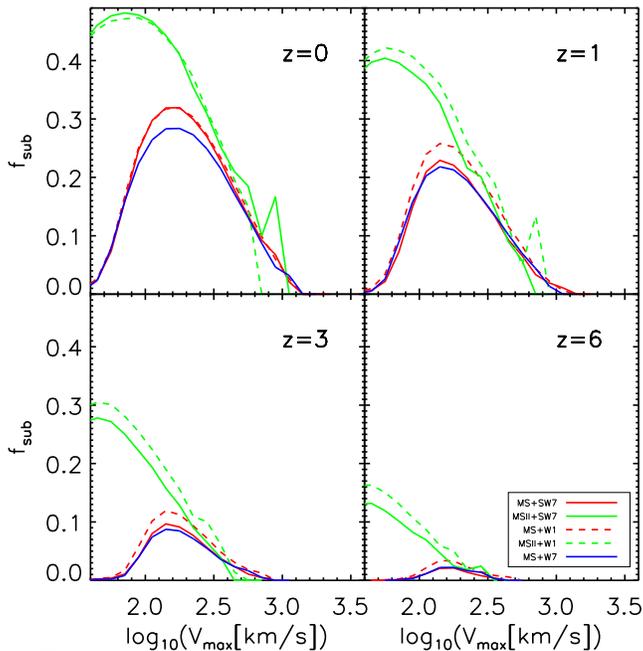}}\\%
\caption{Fraction of satellite subhaloes as a function of $V_{max,infall}$ 
  at $z\approx 0, 1, 3$ and 6. The various curves are colour and line-type coded
  as in Fig. ~\ref{fig:dmmass} and ~\ref{fig:dmvmax}}
\label{fig:dmsatfrac}
\ec
\end{figure}

Fig.~\ref{fig:dmsatfrac} shows the fraction of subhaloes that are
satellites rather than dominant, ``main'' subhaloes as a function of
$V_{max,infall}$ and redshift. The various curves are colour and
line-style coded as in Figures~\ref{fig:dmmass}
and~\ref{fig:dmvmax}. Not surprisingly, the MS-II, with its 125 times
better mass resolution, always finds higher satellite fractions than
the MS. The threshold at which the satellite fractions converge is
higher than the one where the mass functions and $V_{max,infall}$
functions converge. The satellite fraction increases with decreasing
(sub)halo mass, but never exceeds 50\%. This fraction is also a
decreasing function of redshift. At $z=6$ the maximum value is only
about 10\%. Comparing WMAP1 to WMAP7, we find that satellite fractions
are always higher in WMAP1 which should be reflected in a (mild)
enhancement of satellite galaxy abundance in this cosmology. Once again there
is good agreement between the direct simulation of the WMAP7 cosmology
(MS-W7, solid blue curves) and the results from the scaled MS
simulation (solid red curves), although at $z=0$, the direct
simulation predicts a 4\% lower satellite fraction than the rescaled
one at $\log V_{max,infall} = 2.25$. This is probably a consequence of
the rescaling of the time axis which results in an underestimate of
the number of orbital times available for merging in the scaled
simulation.

\begin{figure}
\bc
\hspace{-0.6cm}
\resizebox{8.5cm}{!}{\includegraphics{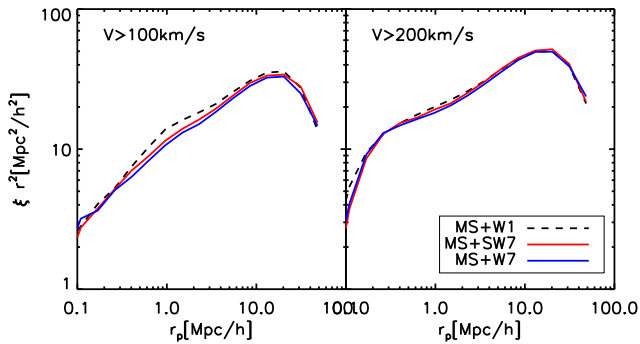}}\\%
\caption{Autocorrelation functions for (sub)haloes with
  $V_{max,infall} >  100$ (left panel)
  and 200~km/s (right panel). Solid red curves are for the MS scaled to the WMAP7
  cosmology while solid blue curves are for MS-W7.  Dashed black
  curves refer to the original MS in its WMAP1 cosmology.}
\label{fig:dmcor}
\ec
\end{figure}

So far we have studied the abundance of satellite and central
(sub)haloes as a function of mass and maximum circular velocity. In
Fig.\ref{fig:dmcor} we compare their spatial clustering by plotting
the two-point correlation function for all (sub)haloes with $V_{max,
infall}$ greater than 100 and 200 km/s, where $V_{max,infall}$ is
defined as in Fig.~\ref{fig:dmvmax}.  The results in this plot are
based on the MS, MS-SW7 and MS-W7 only. Although results for the two
cosmologies are very similar, the correlation function is slightly
higher in WMAP1 than in WMAP7 for both maximum circular velocity
limits. This is consistent with the larger bias expected for given
halo mass at lower $\sigma_8$, and with the larger satellite fractions
in WMAP1. Again the results for the MS scaled to the WMAP7 cosmology
agree very well with the direct measurements in MS-W7.  Given the
small size of the differences between the two cosmologies at all redshifts, we can
anticipate that galaxy formation models will produce similar galaxy
populations in the two cases for similar values of their efficiency
parameters. We will explore this in great detail throughout section 3.

 \subsection{Semi-analytic modelling of galaxy formation}

In the standard galaxy formation scenario, gas and dark matter fall
together into growing dark haloes, both diffusely and in clumps. The
gas then shocks, radiating away its infall energy either immediately or
more slowly from a hot quasi-static atmosphere, and settling into a
central rotationally supported gas disk. As discussed in the original
papers \citep{Rees1977,White1978,White1991}
the rapid cooling or ``cold flow'' regime is dominant at early times
and in lower mass haloes, while the cooling flow regime is more
important at late times and in massive haloes
(see \cite{Voort2011} for a recent discussion). When the gas
disk is sufficiently massive it starts to form stars, as well as to
build a central black hole. The evolving stellar populations then pump
energy, mass and heavy elements into their surroundings through
stellar winds and supernovae, and the accreting black hole also heats
its environment. As satellite galaxies orbit within larger haloes
(``clusters'') they are affected by dynamical processes which can
strip their remaining gas, can tidally truncate their stellar and dark
matter components, and can cause them to merge with the central
galaxy, stimulating starbursts and further AGN activity. These cooling
and feedback processes regulate the growth of galaxies and shape their
mass and luminosity functions.

In this work we use the galaxy formation model of \cite[][hereafter
G11]{Guo2011} to follow all these baryonic processes
within dark matter (sub)halo trees which describe the evolution of
nonlinear structure within the MS, the MS-II and the MS-W7.  This is
the latest version of the simulation-based model of the Munich
group
\citep{Kauffmann1999,Springel2001,Springel2005,Croton2006,Lucia2007}. G11
adjusted model parameters so that galaxy abundances in the MS and
MS-II agreed with those measured in the Sloan Digital Sky Survey
(SDSS) as a function of stellar mass, luminosity, size, star formation
rate, colour, morphology, gas content, metallicity and characteristic
velocity. Although they were able to reproduce many observed
properties of galaxies both in the local universe and at high
redshift \citep[see also,][]{Henriques2012}, G11 highlighted
three significant problems.  At early times ($z\geq 1$) the abundance
of low-mass galaxies is significantly overpredicted. At low redshift,
satellite galaxies in groups and clusters are too uniformly
red \citep[see also,][]{Weinmann2011}. Finally, low-mass
galaxies are too strongly clustered on scales below about 1~Mpc.

These problems may be related: low-redshift, red satellites are the
descendants of low-mass galaxies which formed and fell into larger
systems at early times, thus increasing small-scale clustering. As we
saw above there are fewer massive haloes at high redshift and less
substructure at low redshift in the WMAP7 cosmology than in WMAP1. In
this paper we therefore apply the G11 galaxy formation model to the WMAP7 cosmology to investigate
whether the problems are alleviated.  We follow the same philosophy as
in the original paper, adjusting the efficiency parameters of the
model to fit galaxy abundances in the SDSS as a function of galaxy properties. We then compare the evolution and clustering of
galaxies with other data in order to test the model.  In particular,
in this paper we compare results for the two cosmologies. In the
remainder of this subsection, we briefly describe our treatment of the
processes for which parameters are readjusted in order to maintain
agreement between model predictions and local observations when the
cosmology is changed.

In the G11 model, the star formation rate is
assumed to be proportional to the mass excess above threshold of the
cold gas disk and inversely proportional to its rotation period:
\begin{equation}
\dot{M}_*=\alpha(M_{\rm gas}-M_{\rm crit})/t_{\rm dyn},
\end{equation}
where $\alpha$ is a free parameter describing the star formation
efficiency. Recent studies of galaxy-wide star formation suggest
$\alpha\sim 0.02$ for present-day disk galaxies, although with a
characteristic time which is determined by the atomic-to-molecular
transition rather than the disk orbital
period \citep{Bigiel2011}. We operationally define the
orbital or dynamical time as $t_{\rm dyn} = 3R_{\rm gas,d}/V_{\rm
max}$, where $R_{\rm gas,d}$ is the gas disk scale length and $V_{\rm
max}$ is the maximum circular velocity of the (sub)halo. $M_{\rm crit}$ is a critical
mass above which stars can form, obtained
by integrating the critical surface density assuming a flat rotation
curve and a gas velocity dispersion of 6 {\rm km/s}: 
\begin{equation}
\label{eq:mcrit} 
M_{\rm crit}=11.5\times
10^9\left(\frac{V_{\rm max}}{200{\rm km/s}}\right)\left(\frac{R_{\rm
gas,d}}{10{\rm kpc}}\right)M_{\odot}.
\end{equation}

As stars evolve, 43\% of the total mass of each generation is returned
to the interstellar medium immediately, a crude approximation to the
integrated mass loss which more realistically should be spread out
over a period extending to several Gyr.  Supernova explosions are also
assumed to take place instantaneously, releasing energy to heat surrounding gas, as well as heavy elements to enrich
it. The total energy released by supernovae is modelled as
\begin{equation}
\label{eq:SNeject}
\delta E_{\rm SN} =\epsilon_{\rm halo} \times\frac{1}{2}\delta M_* V_{\rm SN}^2.
\end{equation}
where $0.5V_{\rm SN}^2$ is the mean kinetic energy of supernova ejecta
per unit mass of stars formed. $\epsilon_{\rm halo}$ is a
halo-dependent efficiency:
\begin{equation}
\label{eq:SNeff}
\epsilon_{\rm halo}=\eta\times\left[0.5+\left(\frac{V_{\rm
        max}}{V_{eject}}\right)^{-\beta_1}\right],
\end{equation}
where $\eta$ is an adjustable parameter, $\beta_1$ describes the
dependence on $V_{\rm max}$, and $V_{eject}$ sets the
normalisation. Note that $\epsilon_{\rm halo}$ is assumed to saturate
at unity, i.e., the total energy reheating/ejecting gas from galaxies
cannot exceed the total amount of energy provided by SNe.

Part of this energy is used to reheat gas from the ISM to a hot gas halo:
\begin{equation}
\label{eq:Mreheat}
\delta M_{\rm reheat} = {\rm min}[\epsilon_{\rm disk} \times \delta M_*, 2\delta E_{\rm SN}/V_{\rm max}^2].
\end{equation}
where $\delta M_*$ is the mass of newly formed stars and
$\epsilon_{\rm disk}$ is another halo-dependent efficiency similar to
that of Eq.~\ref{eq:SNeff},
\begin{equation}
\epsilon_{\rm disk}=\epsilon\times\left[0.5+\left(\frac{V_{\rm
        max}}{V_{\rm reheat}}\right)^{-\beta_2}\right].
\end{equation}
Here $\epsilon$, $\beta_2$ and $V_{\rm reheat}$ are again
adjustable parameters. When the first term on the {\it rhs} of Eq.~\ref{eq:Mreheat}
is the smaller, residual feedback energy is used to eject hot gas out
of the halo altogether (see G11).

One of the most frequently invoked mechanisms to quench star formation
in the central galaxies of clusters is feedback from a radio AGN.
Following \cite{Croton2006}, G11 adopt a model in which the
heating rate from radio AGN is expressed as
\begin{eqnarray}
\dot{E}_{\rm AGN} &=& 0.1c^2  \dot{M}_{\rm BH} \nonumber \\
&=& 0.1c^2 \kappa \left(\frac{f_{\rm hot}}{0.1}\right)\left(\frac{V_{\rm vir}}
{200{\rm km/s}}\right)^{3}\left(\frac{M_{\rm
BH}}{10^8/{\rm h}M_{\odot}}\right), 
\end{eqnarray}
where, $M_{\rm BH}$ is the mass of the black hole, $f_{\rm hot}$ is
the hot gas fraction of the halo, $V_{\rm vir}$ is the circular
velocity at $R_{200}$ and $\kappa$ parametrises the efficiency of hot
gas accretion. Massive halos form later in WMAP7 than in WMAP1, so one
expects that a lower value of this accretion efficiency will be
required if central galaxies are to grow to the same mass as in the
WMAP1 cosmology.

In the code module which models the disruption of type 2 galaxies
(satellites which have lost their subhaloes) G11 accidentally failed
to use updated positions. Effectively, this meant that these objects
were either disrupted when they first lost their subhaloes, or not at
all. This bug has been fixed in the current version of the code, which
now correctly identifies the position of each type 2 in this module
(as in the rest of the code) with that of the most bound particle in
its subhalo at the last time this was identified, modified by a
shrinking factor to account for dynamical friction. This correction
results in small differences between the WMAP1 results of this paper
and those in G11.

\section{Results}
\label{sec:result}

In this section, we compare predicted galaxy properties in the WMAP1
and WMAP7 cosmologies. As noted above, we readjust star formation and
feedback parameters for the WMAP7 cosmology so that the low-redshift
stellar mass function for the scaled MS and MS-II simulations matches
that inferred from the SDSS, just as was done for WMAP1
by \cite{Guo2011}.  The galaxy formation parameters we adopt for the
two cosmologies are summarised in Table ~\ref{table:sam} where they
differ. Parameters not listed in this table are held to the values
adopted by G11. Note that there are some degeneracies between
parameters in these models, so that the sets we use here are not the
only ones which can produce fits to the observations of the quality we
show.

Overall, fitting the observed stellar mass function in the WMAP7
cosmology requires somewhat lower star formation and feedback
efficiencies than in WMAP1. Together with the somewhat later formation
of structure, this results in a relatively higher fraction of blue
galaxies in the WMAP7 cosmology. The dependence of SN feedback on
$V_{max}$ that we adopt in WMAP7 is weaker than in WMAP1, resulting in
less efficient ejection by winds from low-mass systems. Since massive
structures form later in WMAP7, less efficient AGN feedback is
required to allow brightest cluster galaxies to grow to the observed
size. We now analyse trends of this kind in more detail.

\subsection{Stellar mass and luminosity functions}
\label{sec:mf}
\begin{figure}
\bc
\hspace{-0.6cm}
\resizebox{8.5cm}{!}{\includegraphics{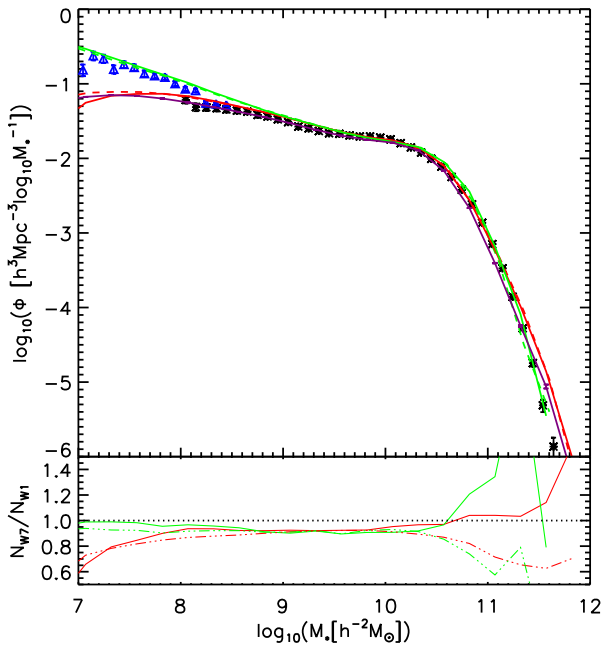}}\\%
\caption{Stellar mass functions predicted by our galaxy formation
  models. Symbols are the observed SDSS mass functions of Li \& White
  (2009) and Baldry et al. (2008). Red and green curves refer to
  models implemented on the MS and MS-II, respectively. Solid curves
  are for the WMAP7 cosmology, while dashed ones are for WMAP1. In the
  bottom panel, solid lines compare the ratio of the galaxy abundances
  in these WMAP7 and WMAP1 models as a function of stellar mass; again
  red is for the MS and green is for the MS-II. The difference is less
  than 8\% except at the very high mass end. {This panel also
  shows as dashed-dotted curves the result of applying the original
  WMAP1 galaxy formation parameters to the scaled WMAP7
  simulations. Even without retuning, the change in cosmology affects
  the mass function only at about the 10\% level (except at the
  highest masses)}. Finally, the upper panel also shows the result of
  applying our WMAP7 galaxy formation model to the MS-W7, a repeat of
  MS carried out directly in the WMAP7 cosmology (purple curve).}
\label{fig:MF}
\ec
\end{figure}
 
Fig.~\ref{fig:MF} compares model stellar mass functions for the WMAP1
(dashed) and WMAP7 (solid) cosmologies to the observational data from
\cite{Li2009} and \cite{Baldry2008} which we have fitted. Red
and green curves refer to the MS and the MS-II, respectively. Our
galaxy formation modelling is able to match the observed galaxy
abundance equally well in the two cosmologies. {Observational
stellar mass estimates should differ slightly in our two cosmologies
due to the geometry dependence of luminosity distance (we already
account for the difference in $H$). This residual difference is about
0.3\% at z=0.1 and reaches 3.7\% at z = 1.5. These effects are small
enough that we neglect them in the following.}

At masses less than $10^{10.5}M_{\odot}$ the predicted abundances are
almost identical in the two cosmologies, as shown by the detailed
comparison in the lower panel. At higher masses, the WMAP7 model
predicts slightly more galaxies. This reflects the specific AGN
feedback efficiency we adopt for this cosmology, which determines the
masses of the biggest galaxies. {Although we re-tuned the galaxy
formation parameters to reproduce the $z=0$ stellar mass function in
the WMAP7 cosmology, the difference in mass function evolution between
the two cosmologies is, in fact, sufficiently small that a good fit is
obtained even if we do not change these parameters at all. This is
illustrated in the lower panel of Fig.~\ref{fig:MF} where the
dashed-dotted curves show the ratio of the mass functions obtained
when the WMAP1 galaxy formation model is run on the scaled WMAP7
simulations to those obtained when it is run on the original WMAP1
simulations. These mass functions are within 10\% of eacxh other
except at the highest masses.}

To explore how well our rescaling technique works, we have also
implemented our WMAP7 galaxy formation model with unchanged parameters
on a repeat of the MS carried out directly using the updated WMAP7
cosmology (MS-W7).  We overplot the resulting stellar mass function as
a purple curve in the upper panel of Fig.~\ref{fig:MF}. The results
are very similar, particularly when account is taken of the fact that
the effective mass resolution of the scaled MS simulation is {poorer} by a factor of 1.23 than that of MS-W7, and that the initial
phases of the simulations are different so that the abundance
comparison is affected by cosmic variance \citep{Smith2012}.  The two
functions differ primarily through a small offset of up to 0.1dex in
stellar mass, with the direct simulation producing systematically
slightly lower mass galaxies than the scaled MS. This offset leads to
very small abundance differences at stellar masses below that of the
Milky Way, but to differences approaching a factor of 2 at the highest
masses where cosmic variance is substantial. For the rest of this
paper, we will compare the scaled and unscaled Millennium Simulations
in order to be able to test for numerical convergence by comparing MS
and MS-II and to avoid cosmic variance effects at high mass.

\begin{figure}
\bc
\hspace{-0.6cm}
\resizebox{8.5cm}{!}{\includegraphics{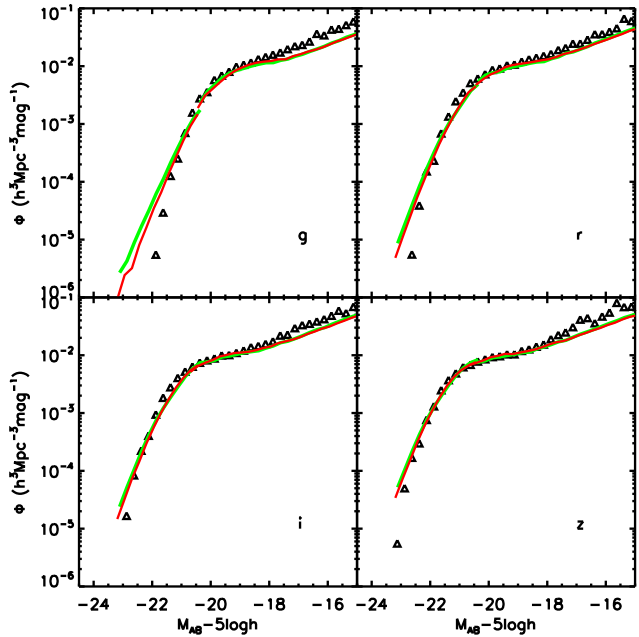}}\\%
\caption{Galaxy luminosity functions in the SDSS $g$, $r$, $i$ and $z$
photometric bands. The smooth green and red curves are predictions from our WMAP7 
and WMAP1 models, respectively and are taken from the MS at the bright end and from the MS-II
at absolute magnitudes below about $-20.0$. The symbols are observational data
for a low-redshift SDSS sample taken from Blanton et al. (2005).}
\label{fig:LF}
\ec
\end{figure}

Fig.~\ref{fig:LF} shows luminosity functions in the SDSS $g$, $r$, $i$
and $z$ bands. Symbols are taken from Blanton et al. (2005); green and
red curves are model predictions for the WMAP7 and WMAP1 cosmologies,
respectively. We include the $h$ factor in the magnitude and volume
units in order to compare the observational data to the different
cosmologies in a fair way. Model predictions are from the MS for
galaxies with absolute magnitudes brighter than -20 and from the MS-II
for fainter galaxies. For both cosmologies the predictions agree
reasonably well with observation in all four bands. In the $r$ band,
the predictions are identical in two cosmologies over the full
magnitude range. For very bright galaxies, the $g$-band luminosity
function is higher in WMAP7 than in WMAP1 which already overpredicts
the observed abundance. As pointed out by G11 this may
reflect the inadequacy of our dust model in massive star-bursting
galaxies. WMAP7 gives slightly higher results since it predicts
slightly more major mergers of gas-rich massive galaxies at low
redshift.

\subsection{Stellar mass vs. Infall halo mass relation}
\label{sec:mstar2mhalo}

\begin{figure}
\bc
\hspace{-0.6cm}
\resizebox{8.5cm}{!}{\includegraphics{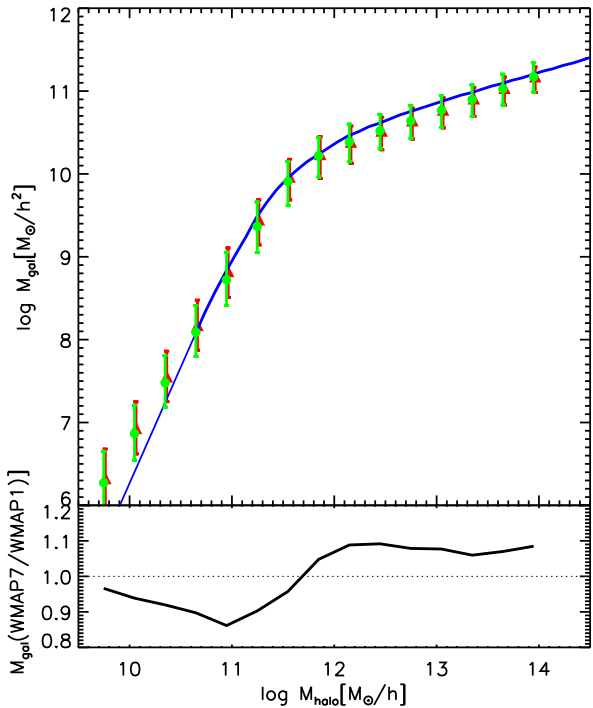}}\\%
\caption{Galaxy stellar mass as a function of maximum past halo mass,
  as predicted for the WMAP7 (green) and WMAP1 (red)
  cosmologies. Symbols with error bars represent the median values and
  the $\pm 1 \sigma$ scatter in the models. They are offset slightly
  in the $x$-direction for clarity.  The blue curve is the WMAP1
  relation derived directly from the SDSS stellar mass function and
  from subhalo abundances in the MS and MS-II under the assumption
  that the two quantities are monotonically related without scatter
  (Guo et al. 2010). The lower panel shows the ratio of the central
  valus of the red and green bars.}
\label{fig:gf}
\ec
\end{figure}

Galaxies form by the condensation of gas at the centres of dark matter
haloes, and as a result there is a relatively tight relation between
their stellar masses and the dark matter masses of their haloes.
After a galaxy falls into a more massive system, its halo mass can be
reduced substantially by tides, while the stellar distribution is much
less affected.  For such satellites, stellar mass is more closely
related to halo mass just before infall than to current halo mass.
Thus, many authors have assigned galaxies to (sub)haloes in dark
matter simulations by assuming a monotonic relation between stellar
mass and maximum past (sub)halo mass and forcing the simulation to
reproduce the observed abundance of galaxies as a function of stellar
mass \citep[e.g.][]{Vale2004,Conroy2006,Moster2010,Guo2010}.  For a concordance $\Lambda$CDM cosmology, the
relation obtained through such subhalo abundance matching is
consistent with observational estimates from weak lensing and
satellite galaxy dynamics, and results in galaxy correlations as a
function of stellar mass which are in quite good agreement with
observation.  Nevertheless, galaxy properties depend also on halo
properties other than mass. For example, at given halo mass, more
concentrated haloes form earlier and are denser; as a result they form
stars more efficiently.  Direct estimation of such effects from
semi-analytic models can help to understand the scatter in the
$M_\star$-$M_{halo}$ relation.

Together, the MS and the MS-II provide sufficient resolution and
statistics to measure the abundance of dark matter subhaloes across
seven orders of magnitude in infall mass (Fig.~\ref{fig:dmmass}).
Since this mass function varies very little between WMAP7 and WMAP1,
and our galaxy formation parameters are chosen to reproduce the
observed stellar mass function in both cases, the relation between
stellar mass and infall halo mass barely changes between the two
cosmologies. In the upper panel of Fig.~\ref{fig:gf}, error bars are
centred on the median value and have length equal to twice the {\it
rms} scatter in stellar mass for given infall (sub)halo mass and are
shown in green for WMAP7 and in red for WMAP1.  The ratio of the two
median stellar masses is shown in the lower panel.  A blue line gives
the relation obtained by \cite{Guo2010} for the WMAP1 case when
assuming a monotonic scatter-free relation and forcing the stellar
mass function to follow the observational data of Fig.~\ref{fig:MF}
exactly. As expected, both simulation results are consistent with the
abundance matching relation over the mass range $10^{10} -
10^{14}M_{\odot}/h$. At low mass, galaxies of given stellar mass reside
in more massive halos in WMAP7, while at high mass they are hosted by
less massive halos. As can be seen from the ratio plot in the lower
panel, above a halo mass of about $5\times10^{11}M_{\odot}/h$, galaxy
formation is more efficient in WMAP7 than in WMAP1, while in lower
mass haloes it is the other way around.  The effects are very small,
however:  galaxy masses at  fixed halo mass vary by a maximum of about 10\%.

\subsection{Gas-phase metallicities}
\begin{figure}
\bc
\hspace{-0.6cm}
\resizebox{8.5cm}{!}{\includegraphics{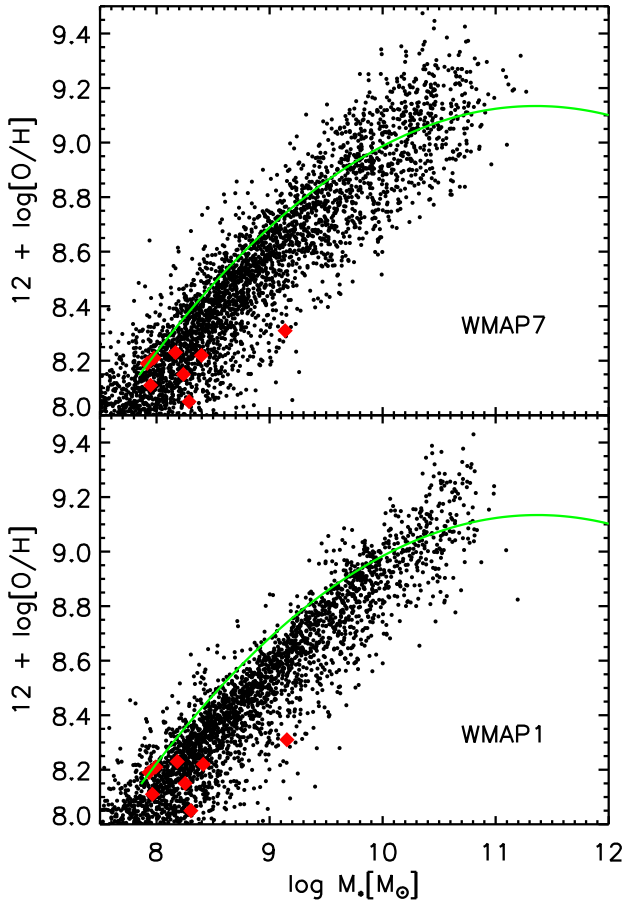}}\\%
\caption{Cold gas metallicity as a function of stellar mass. The top panel 
shows results for WMAP7, and the bottom panel for WMAP1.  In both
panels, the solid curves represent observational results for the SDSS
from Tremonti et al. (2004), while red diamonds are taken from Lee et
al. (2006).}
\label{fig:metals}
\ec
\end{figure}

Fig.~\ref{fig:metals} plots gas-phase metallicity against stellar mass
for star-forming galaxies, defined to be those with specific star
formation rate higher than $10^{-11}$/yr. The upper panel is for the
WMAP7 and the lower panel for WMAP1. In each case black dots are
randomly selected simulated galaxies from the MS-II, while the curves
and the red diamonds represent observational data \citep{Tremonti2004, Lee2006}. Predictions in both cosmologies agree with
the observations reasonably well (the same yields were assumed in the
two cases). The median metallicity is about 0.1 dex higher in WMAP7 than in
WMAP1, thus closer to the observations, and the scatter is slightly
bigger. In neither cosmology does the model reproduce the observed
turnover in gas-phase metallicity at the highest masses.

\subsection{$u-i$ colour distribution}
\begin{figure}
\bc
\hspace{-0.6cm}
\resizebox{8.5cm}{!}{\includegraphics{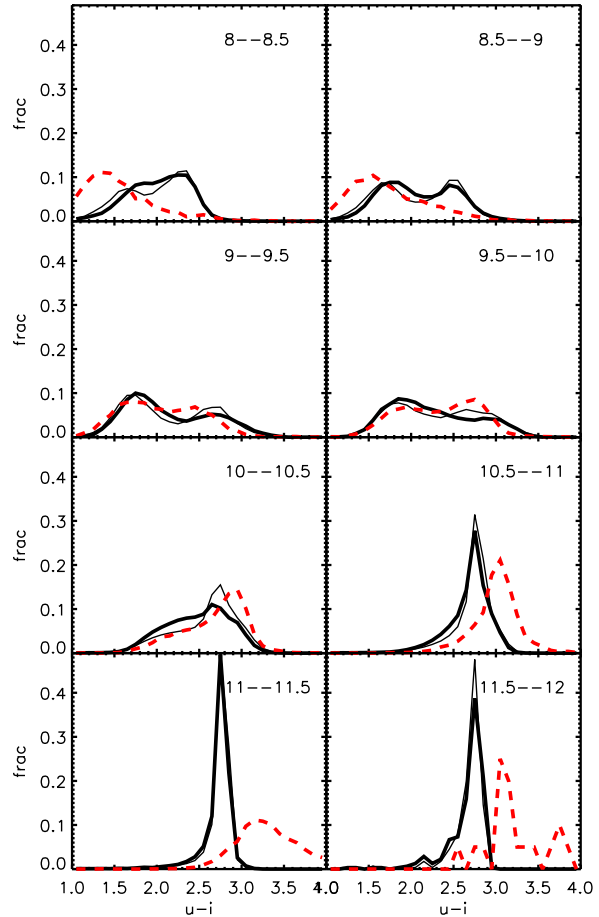}}\\%
\caption{$u-i$ colour distributions as a function of stellar mass. Thick solid black
curves show the distributions predicted by our preferred WMAP7 model,
while thin curves are for WMAP1.  Dashed red curves are distributions
compiled from SDSS/DR7. }
\label{fig:colorui}
\ec
\end{figure}
\begin{figure}
\bc
\hspace{-0.6cm}
\resizebox{8.5cm}{!}{\includegraphics{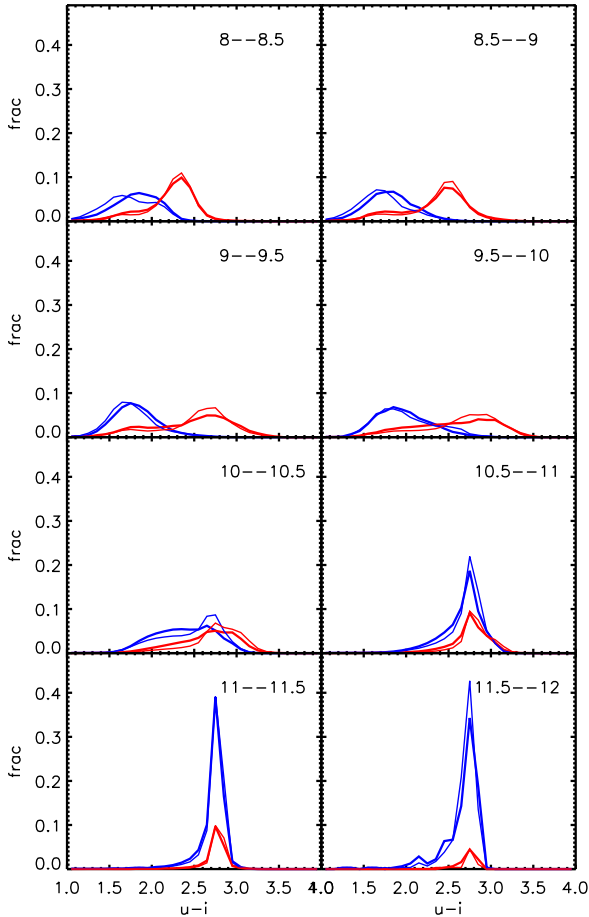}}\\%
\caption{$u-i$ colour distributions as a function of stellar mass for
centrals (blue) and satellites (red). Thick curves are results for
WMAP7 while thin curves are for WMAP1.}
\label{fig:colorui_type}
\ec
\end{figure}

One of the main problems found by G11 with their WMAP1 model was that
it predicts dwarf galaxies that are too red. This may be an indication
that dwarfs form too early, and, indeed, at $z>0.8$ the model
substantially overpredicts the abundance of galaxies less massive than
a few $10^{10}M_\odot$. Structures form somewhat later in the WMAP7
cosmology, so one might expect a delay in the formation of galaxies,
leading to a bluer $z=0$ colour at given stellar mass.
Fig.~\ref{fig:colorui} compares model colour distributions for WMAP7
(thick black lines) and WMAP1 (thin black lines) with SDSS data (red
dashed lines) as a function of stellar mass. In general, the WMAP7
model does predict slightly bluer galaxies than WMAP1, but the effect
is small and the disagreement with the observations remains large,
particularly for low-mass galaxies.

In Fig.~\ref{fig:colorui_type} we investigate this problem further by
separating model galaxies into central (blue) and satellite (red)
populations.  Results for the two cosmologies are almost identical. In
both cases, almost all the red population at low stellar mass is
contributed by satellites \citep[see also,][]{Weinmann2011}. Above
about $10^{9}M_{\odot}$, the predicted red fraction is consistent with
observation, but at lower masses the models show a peak of red
satellites which is missing in the SDSS data.  This could either be
because too many low-mass galaxies form at high redshift and survive
as satellites to the present day, or be because star formation is
quenched too efficiently in satellites in the simulations. Another
possibility is that the red satellite population has been missed in
the SDSS data, perhaps because its surface brightness is too low.

\subsection{Cluster number density profiles}
\begin{figure}
\bc
\hspace{-0.4cm}
\resizebox{8.5cm}{!}{\includegraphics{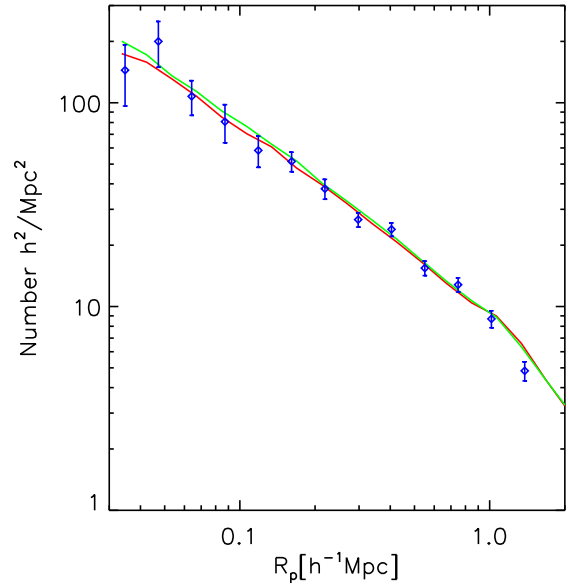}}\\%
\caption{Projected number density profiles for galaxies with stellar
  mass above $6\times10^{9}h^{-2}M_\odot$ in clusters with $10^{14}M_\odot <
  M_{200}<2\times 10^{14}M_\odot$. The green curve is the result for
  WMAP7 while the red curve is for WMAP1. Blue symbols with error
  bars are from the SDSS (see G11 for further details). }
\label{fig:cluster}
\ec
\end{figure}

G11 showed that their WMAP1 model was able to reproduce the observed
mean number density profile of galaxies in rich galaxy clusters
between about 30~kpc and 1.5~Mpc both in the MS and in the MS-II. They
used this to argue that their treatment of ``orphan galaxies''
(satellite galaxies for which the corresponding dark matter subhalo
had been completely destroyed by tidal effects) was realistic since
such galaxies are much less important at the higher resolution of the
MS-II. For both simulations they found a projected cluster number
density which appeared marginally higher than observed below about
150~kpc.  Here we revisit this topic, comparing predictions for
cluster structure in the WMAP1 and WMAP7 cosmologies.

We follow the procedures in G11 exactly to make this comparison. We
use a simple 'observational' cluster finder to select galaxy clusters
in the same way in the simulations and in the SDSS. Here we summarize
the main cluster selection criteria; details are given in
G11. Galaxies with mass greater than $6\times10^{9}h^{-2}M_{\odot}$ are
counted around potential BCGs at projected separations $r_p < 1.1{\rm Mpc}/h$
and redshift differences $|\Delta v| < 1200$ km/s. Clusters with
counts in the range $45<N_{g}<105$ are accepted.  In the MS this leads
to a sample of 2295 clusters for the WMAP1 cosmology, and 2443 for
WMAP7. G11 estimate the observed number density of clusters defined in this
way to be $2\times 10^{-6} (h/{\rm Mpc})^3$, closer to the
simulation value for WMAP7 than for WMAP1, although both are probably
consistent within the overall uncertainties. The number of clusters in
the SDSS spectroscopic database which are complete to this stellar
mass limit and satisfy the count criteria is just 31.

Fig.\ref{fig:cluster} compares mean projected number density profiles
for stacks of MS clusters in each of the two cosmologies and in the
SDSS data. The green curve is for WMAP7, the red for WMAP1 and the
blue symbols for SDSS, with error bars indicating the {\it rms}
cluster-to-cluster scatter. The overall agreement between simulation
and observation is quite good with the model perhaps overproducing
galaxies close to cluster center.  Results for the two cosmologies
are almost identical with WMAP7 (green curve) giving a very slightly
higher amplitude, consistent with the higher galaxy formation
efficiency for WMAP7 in this mass range (see Fig.~\ref{fig:gf}).

\subsection{Correlation Functions}
\label{sec:correl}

\begin{figure*}
\bc
\hspace{-0.6cm}
\resizebox{12cm}{!}{\includegraphics{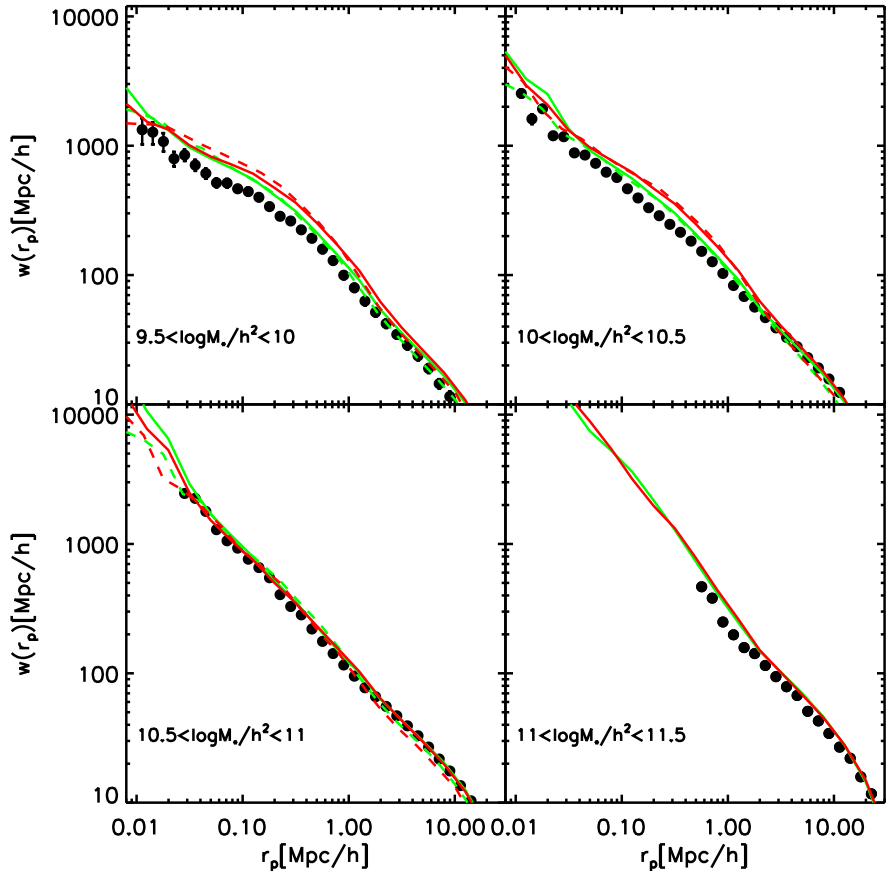}}\\%
\caption{Projected autocorrelation functions for galaxies in different 
stellar mass ranges. Symbols with error bars are results for SDSS/DR7
calculated using the techniques of Li et al. (2006). Solid and dashed
curves give results for our preferred model applied to the MS and the
MS-II, respectively. Green curves are for the WMAP7 and red for WMAP1.
In general the WMAP7 results are slightly lower than WMAP1, particularly
for lower mass galaxies and on small scales where the one-halo term due
to satellite galaxies is dominant.}
\label{fig:cor}
\ec
\end{figure*}

In G11 the parameters of the galaxy formation model were tuned to
reproduce the observed abundances of galaxies, particularly their
stellar mass functions and the distributions of internal properties
like colour, morphology and gas content. The clustering and the
evolution of the population were then used as tests of the model. For
massive galaxies the observed low-redshift autocorrelation functions
were well reproduced from scales of 20~kpc out to $\sim 30$~Mpc. For
less massive galaxies the agreement remains good on large scales where
correlations are due to pairs of galaxies inhabiting different dark
matter haloes, but the clustering is noticeably too strong on smaller
scales where pairs are predominantly members of the same halo. G11
suggested that this discrepancy might reflect the fact that the
overall amplitude of dark matter structure, in particular the
parameter $\sigma_8$, is higher in the WMAP1 cosmology than estimated
from more recent data.

In Fig. ~\ref{fig:cor} we compare projected two-point correlation
functions for our WMAP1 and WMAP7 models to observational data from
SDSS/DR7. The observational data here are exactly as in G11, but the
WMAP1 model predictions have changed slightly because of the
correction of minor software bugs in the galaxy disruption module (see
Sec. \ref{sec:sim}). For each cosmology we show results for both the
MS (solid lines) and the MS-II (dashed lines). It is reassuring that
the agreement between the two simulations is very good in both
cosmologies despite the two orders of magnitude difference in their
mass resolution. This gives us confidence that our clustering results
are well converged even for low-mass galaxies and on small scales.  On
large scales and for high-mass galaxies there is almost no difference
in clustering between the two cosmologies. However, for pair
separations below $\sim 1$~Mpc and stellar masses below $\sim
10^{10.5}M_\odot$, the correlations are clearly weaker for WMAP7
(green) than for WMAP1 (red), although they remain somewhat higher
than observed in SDSS.  Notice that the difference in large-scale
clustering between the two cosmologies is considerably smaller for the
galaxies than for the mass. This is because the difference in
$\sigma_8$ is largely compensated by a difference in bias, which is
significantly higher at given stellar mass in WMAP7 than in WMAP1.
{Note that for the rescaled WMAP7 cosmology, we do not adjust
large-scale modes according to the final step of the \cite{Angulo2010}
procedure. Tests show that this causes only very small changes in MS
correlations on the scales we analyse here, and it has no measurable
effect on the MS-II correlations.}
\subsection{Results at higher redshift}

So far we have compared our predictions for galaxy formation in the WMAP1 and WMAP7 cosmologies to low redshift data, using the
observed galaxy abundance as a function of stellar mass, luminosity,
metallicity and gas content to set our adjustable efficiency
parameters, and then searching for dependencies on background cosmology
using clustering and colour data. Here we compare predictions for the
two models to higher redshift data in order to test whether an updated
cosmology can solve the evolutionary problem identified by G11:
although their WMAP1 model reproduced the observed cosmic star
formation history moderately well and was consistent with the observed
abundance of massive galaxies out to $z\sim 4$, it overpredicted the
abundance of lower mass galaxies ($M_\star < 10^{10.5}M_\odot$) at
$z\geq 1$. Dwarf galaxies seem to form too early and to age too
quickly in this model.
\begin{figure}
\bc
\hspace{-0.6cm}
\resizebox{8.5cm}{!}{\includegraphics{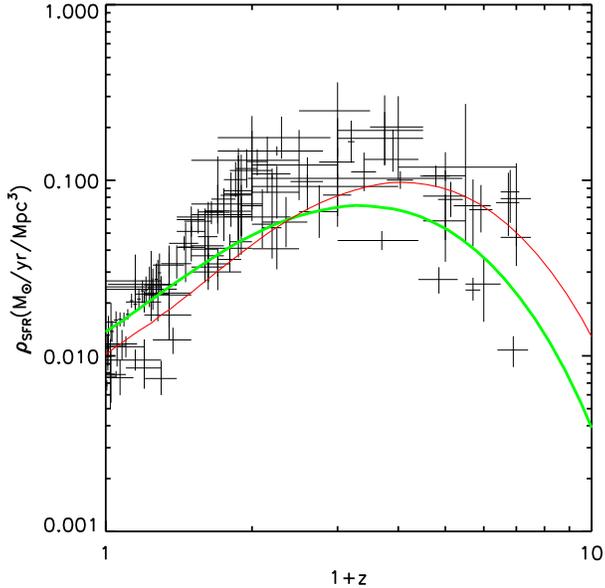}}\\%
\caption{Cosmic star formation rate density as a function of redshift. The
crosses are individual observational estimates compiled by Hopkins et
al. (2007).  The thick solid green curve is obtained from our
preferred WMAP7 model whereas the thin red curve is for WMAP1.}
\label{fig:madau}
\ec
\end{figure}

Fig.~\ref{fig:madau} shows the evolution of the cosmic star formation
rate density.  The green curve is the mean total star formation rate
per unit volume as a function of redshift for our WMAP7 model, while
the red curve is for WMAP1. The symbols are observational estimates
compiled by \cite{Hopkins2007}. In both cases, the predictions of
the models agree moderately well with the observations. Star formation
peaks at significantly lower redshift and the present day star
formation rate density is slightly higher in the WMAP7 model than for
WMAP1. The first of these shifts improves the agreement with the
observational data, while the second makes it worse, particularly
after accounting for the normalisation difference caused by the fact
that integrating a smooth representation of the observational points
in this plot produces a stellar mass density at $z=0$ which is
significantly higher than the value implied by the observed
low-redshift stellar mass function used to set our model parameters.
The later formation of galaxies in WMAP7 is responsible for the bluer
$z=0$ colours visible in Fig.~\ref{fig:colorui}.

\begin{figure*}
\bc
\hspace{-0.6cm}
\resizebox{12.5cm}{!}{\includegraphics{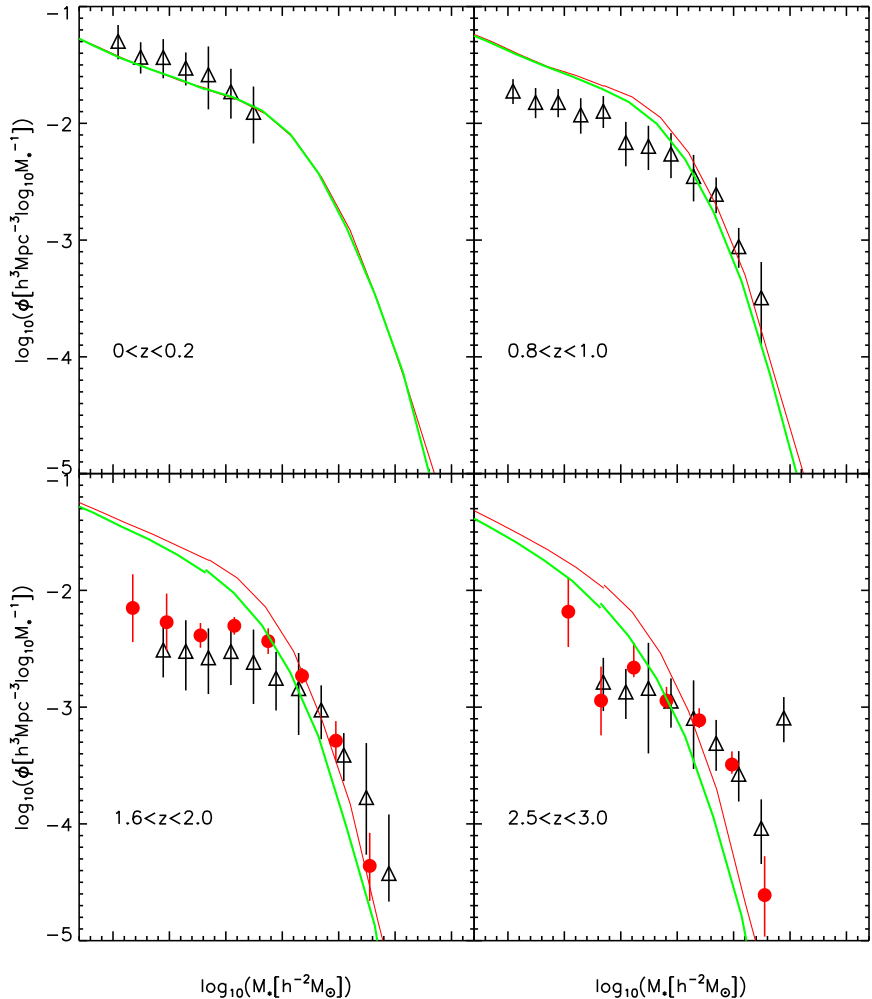}}\\%
\caption{Stellar mass functions for a series of redshift intervals indicated
by the labels in each panel. Observational data are taken from
Perez-Gonzalez et al. (2008) and from Marchesini et al. (2009). The mass
scales of these observational results have been shifted to correct
approximately to the Chabrier IMF assumed in our modelling. Solid
curves are the functions measured from the combination of the MS and
the MS-II for our preferred galaxy formation model, convolving with a
Gaussian of dispersion 0.25~dex in $\log M_*$ to represent the
uncertainty in observational estimates of $M_*$. Green curves are for
our WMAP7 model, while red curves are for WMAP1.}
\label{fig:hzmf}
\ec
\end{figure*}
In Fig.~\ref{fig:hzmf} we investigate how changing cosmology affects
the problem of overly early dwarf galaxy formation identified by G11.
This plot shows stellar mass functions averaged over four disjoint
redshift intervals, as noted in each panel.  The observational data
are the same as used by G11. The red curves, representing our WMAP1
model, are essentially identical to those in their Fig.~23.  It is
clear from the green WMAP7 curves that the later formation of
structure in the newer cosmology has only a minor effect on these mass
functions and does little to reconcile model predictions with
observation. This is not surprising given the similar (sub)halo mass
functions out to at least $z=1$ (see Fig.~\ref{fig:dmmass}). The
discrepancy at low mass must reflect a deficiency in the galaxy
formation physics, rather than in cosmological parameters. Clearly,
star formation at early times must be less efficient in low-mass halos
than the current models assume. This must be compensated by higher
efficiencies at later times so that the $z=0$ stellar mass function is
nevertheless reproduced.

\begin{figure*}
\bc
\hspace{-0.6cm}
\resizebox{12cm}{!}{\includegraphics{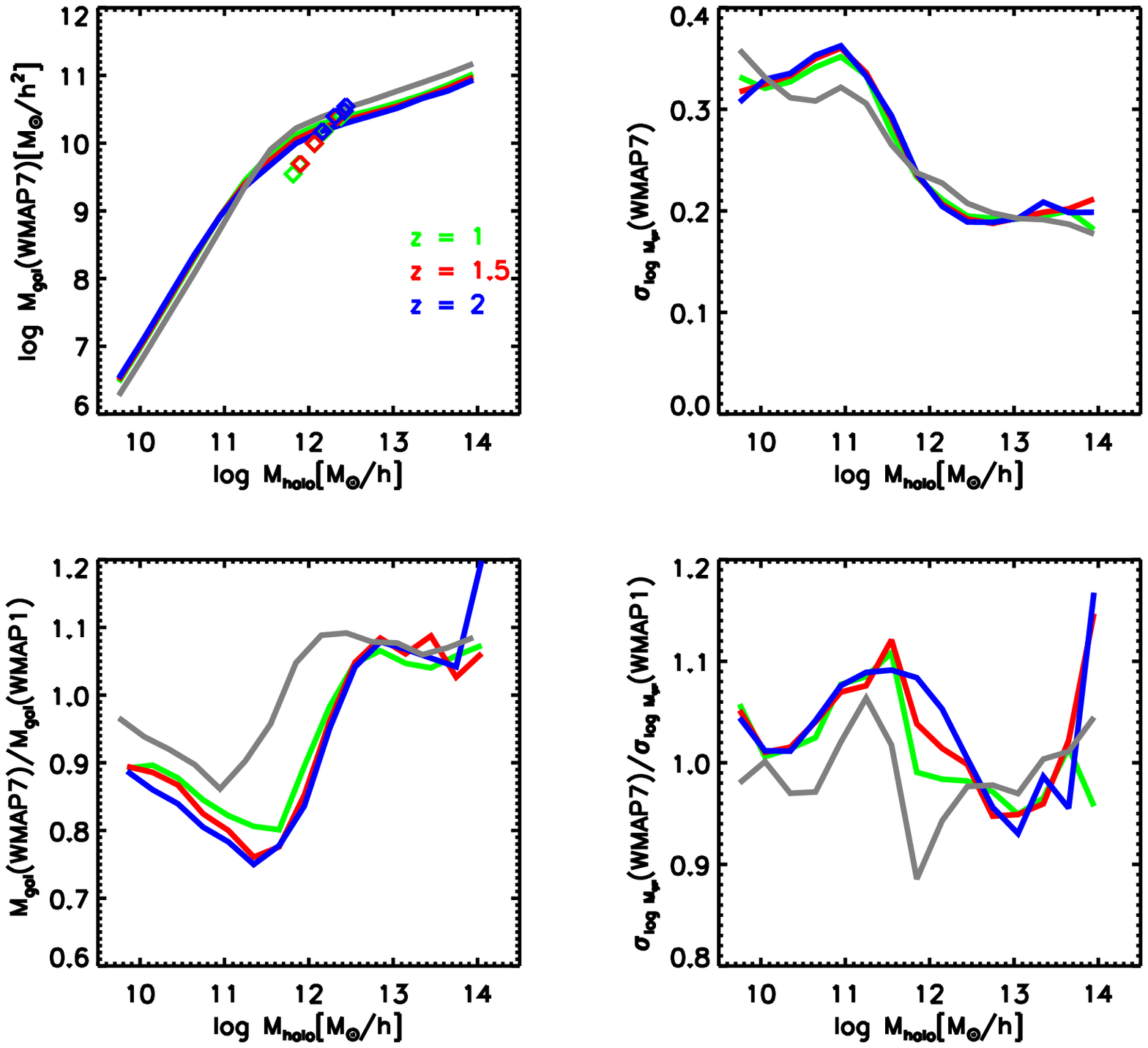}}\\%
\caption{Upper left panel: galaxy stellar mass as a function of maximum past halo mass
  at various redshifts for our WMAP7 galaxy formation model. Symbols
  are observational estimates by Wake et al. (2011) based on fits of
  an HOD model to abundance and clustering data {\it assuming} this
  cosmology. Different colours refer to different redshifts as
  indicated in the bottom right corner. The curves almost coincide and
  are also very close to the $z=0$ curve shown in
  Fig.~\ref{fig:gf}. Bottom left panel: the ratio of galaxy stellar
  mass for WMAP7 to that for WMAP1 as a function of halo mass and at
  various redshifts.  Lines are colour coded as in the upper left
  panel. Upper right panel: one sigma scatter in galaxy stellar mass
  as a function of halo mass. Again, the solid curve is for our WMAP7
  model while symbols show the deviation of observational points of
  Wake et al.(2011) from our WMAP7 model. Right bottom panel: ratio of
  the one sigma scatter in stellar mass between WMAP7 and WMAP1 as a
  function of halo mass.}
\label{fig:gfhz}
\ec
\end{figure*}

As discussed in Sec. ~\ref{sec:mstar2mhalo}, it is instructive to
study the relation between the stellar mass of a galaxy and the
maximum past mass of the halo which hosts it. Here we investigate how
this relations evolves with redshift in our models. The upper left
panel of Fig.~\ref{fig:gfhz} shows the mean relation at redshifts 1.0,
1.5 and 2. The colored curves are for our WMAP7 model and are almost
coincident with each other, but are shifted noticeably with respect to
the $z=0$ curve of Fig.~\ref{fig:gf}, repeated here as a grey curve. The model relation is almost independent of
redshift beyond $z=1$, although at the highest masses, a slightly
lower galaxy formation efficiency is found at earlier times. \cite{Wake2011} use data from the NEWFIRM Medium Band Survey to study the
stellar mass vs. halo mass relation within the same redshift
intervals. Their results, shown with symbols, are based on fitting a
WMAP7 Halo Occupation Distribution model to the abundance and
clustering of stellar mass limited samples of galaxies. Since the
measured abundances are much more constraining than the clustering,
the information content of these points is very similar to that of the
stellar mass function of Fig.~\ref{fig:hzmf}. \cite{Wake2011} found
no significant change in their HOD parameters over $1<z<2$ but a
noticeable shift from $z=0$. The first result is directly visible in
Fig.~\ref{fig:gfhz}, while the second is responsible for the steepness
of the observed relation over the limited stellar mass range for which
it can be estimated, reflecting the lowered amplitude of the stellar
mass function below its characteristic mass which is seen in the data
but not in our models (see Fig.~\ref{fig:hzmf}).

The cosmology dependence of this relation is illustrated in the bottom
left panel of Fig.~\ref{fig:gfhz} which, for galaxies of given maximum
past halo mass, gives the ratio of mean
stellar mass in the WMAP7 and WMAP1 models. Galaxy formation efficiencies are similar in the two
models at all redshifts shown. There is a characteristic halo mass around
$10^{12}M_{\odot}/h$ where the formation efficiency switches from
being higher in WMAP1 to being higher in WMAP7. This mass is closely
related to that where galaxy formation efficiency peaks \cite[e.g.][]{Guo2010} and is higher (and almost constant) for $1\leq z\leq 2$ than
for $z=0$. This shift reflects the ``halo down-sizing'' pointed out by
\cite{Wake2011}. Overall, however, only minor differences in galaxy
formation efficiency are expected between our two cosmologies at
these redshifts.

The upper right panel of Fig.~\ref{fig:gfhz} shows the scatter in
central galaxy stellar mass as a function of maximum past halo mass
for our WMAP7 model. This scatter increases slightly for halo masses
between $10^{9.5}M_{\odot}/h$ and $10^{11}M_{\odot}/h$, reaching
a maximum of $\sim 0.35$~dex before dropping rapidly to 0.2~dex at
masses above $2\times10^{12}M_{\odot}/h$. Again, the scatter does not
depend on redshift and is similar to that at $z=0$.
Notice that it is larger than the differences between the ``observed''
and predicted mean relations in the upper left panel.  Finally, the
lower right panel shows how the predicted scatter changes between our
two cosmologies. Here too differences are small. Below
$10^{12}M_{\odot}/h$ and also at the highest masses the scatter is
higher for WMAP7.

\begin{figure*}
\bc
\hspace{-0.6cm}
\resizebox{12cm}{!}{\includegraphics{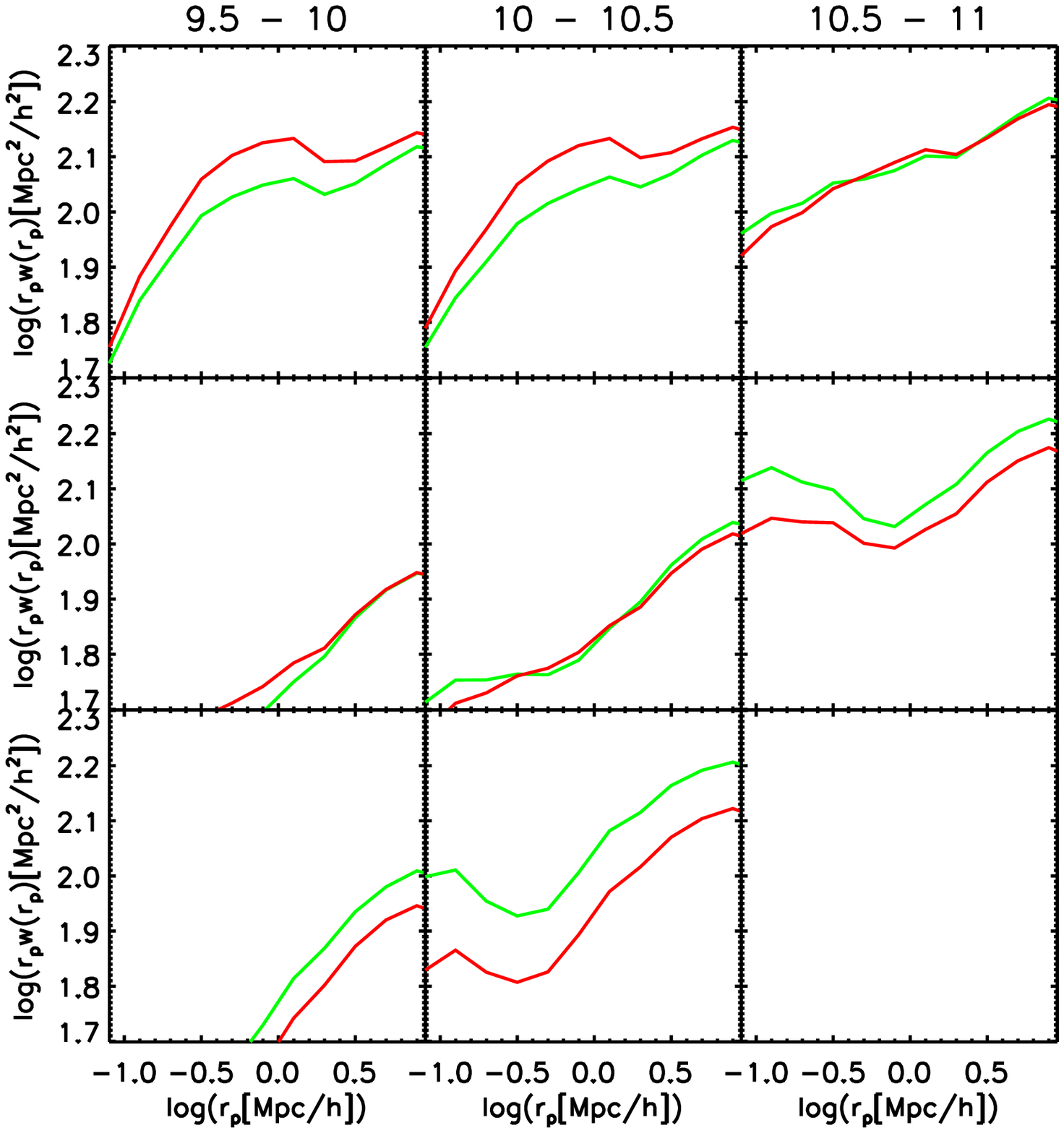}}\\%
\caption{Projected correlation functions (multiplied by $r_p$ as a function of redshift and stellar
  mass for our two cosmologies. Red curves refer to WMAP1 and green to
  WMAP7.}
\label{fig:corhighz}
\ec
\end{figure*}

We finish this section by studying the sensitivity of high-redshift
galaxy clustering to background cosmology. In Fig.~\ref{fig:corhighz}
we show projected correlation functions for three of the stellar mass
bins of Fig.~\ref{fig:cor} and for redshifts 0, 1 and 3. Red curves
are for WMAP1 and green for WMAP7. We have multiplied these functions
by $r_p$ in order to make differences more visible (compare the $z=0$
curves in Figs~\ref{fig:cor} and~\ref{fig:corhighz}). In the local
universe, correlations are predicted to be weaker in WMAP7 than in
WMAP1 for stellar masses below $10^{10.5}M_{\odot}$, and to be very
similar at higher mass. By redshift 1 the predicted correlation
amplitude in WMAP7 has increased substantially relative to WMAP1. The
two now coincide for $M_*<10^{10.5}M_{\odot}$ and WMAP7 is more
clustered at higher stellar mass.  This trend increases with
increasing redshift so that by $z=3$ the WMAP7 model predicts larger
clustering amplitudes than WMAP1 at all stellar masses.  Thus although
the dark matter is substantially less clustered at high redshift for
the more recent cosmological parameters (see Fig.~1), bias effects
cause the opposite to be true for galaxies of any given stellar
mass. The differences are nevertheless quite small over this redshift
range, and current data are insufficiently precise to distinguish the
two cosmologies.

\begin{table*}
\caption{Summary of those parameters of our preferred model which were adjusted
to fit low-redshift observational data, primarily the stellar mass function.}

\begin{tabular}{||l||l||c||c||} 

\hline
Parameter & Description  & WMAP1 & WMAP7  \\
\hline
 $\alpha$       & Star formation efficiency  & 0.02 & 0.011\\
 $\epsilon$     & Amplitude of SN reheating efficiency & 6.5 & 4 \\
 $\beta_1$      & Slope of SN reheating efficiency & 3.5 & 3.2\\
 $V_{reheat}$  &normalization of SN reheating efficiency dependence on Vmax & 70 & 80 \\
  $\eta$     & Amplitude of SN ejection efficiency & 0.32 & 0.18\\
 $\beta_2$      & Slope of SN ejection efficiency & 3.5 & 3.2 \\
 $V_{eject}$  &normalization of SN ejection efficiency dependence on Vmax & 70 & 90\\
 {$\it \kappa$} & Hot gas accretion efficiency onto black holes & 1.5 $\times$ 10$^{-5}$ & 7 $\times$ 10$^{-6}$\\
\hline
\end{tabular} 
\label{table:sam}
\end{table*}

\section{Conclusion}
\label{sec:conclusion}
We have used the re-scaling technique recently developed by Angulo \&
White (2010) to scale the Millennium and Millennium-II simulations from
the WMAP1 cosmology in which they were carried out to the currently
favoured WMAP7 cosmology. The amplitude parameter $\sigma_8$ is lower
but the matter density $\Omega_m$ is higher for WMAP7 than for
WMAP1. These two changes have partially compensating effects,
producing halo mass functions which are similar in the two cosmologies
out to redshifts of at least 3.  As a result, a slightly updated
version of the galaxy formation model of Guo et al. (2011) produces
equally good fits to low-redshift data in the two cases with only
minor adjustments of its parameters.  Furthermore, the predicted
evolution since redshift 3 is also quite similar, although some
residual cosmological dependencies remain, particularly in the
clustering properties.

The main parameters affecting the stellar mass function are the
efficiencies of star formation, of SN and AGN feedback, and of
reincorporation of ejected material. In general, we require lower
star-formation efficiency and weaker feedback in WMAP7 than in WMAP1
in order to produce a similar $z=0$ galaxy population. As a result,
the cosmic star-formation rate peaks later in WMAP7 than in WMAP1, and
galaxies are bluer at low redshift. For halos more massive than
$10^{11.5}M_{\odot}$, galaxy formation is more efficient in WMAP7 than
in WMAP1, while the reverse is true for lower mass haloes.  The
predicted two-point correlations of low-mass galaxies agree better
with observation for WMAP7 than for WMAP1, although their amplitude on
small scales is still slightly higher than observed. The overly early
formation of such galaxies noted by Guo et al. (2011) (see also
\cite{Henriques2012}) is scarcely
affected by the shift to a WMAP7 cosmology. Clearly, observations
require the galaxy formation efficiency to be lowered in low-mass
haloes at early times, and then increased at late times to produce a
similar $z=0$ galaxy population.

Our results contrast with those of \cite{Wang2008} who compared galaxy
properties in simulations of the WMAP1 and WMAP3 cosmologies. For
similar $z=0$ galaxy populations, they found the abundance and
clustering of high-redshift galaxies to differ substantially between
the two cosmologies.  This reflects the fact that their WMAP3 model
assumed $\sigma_8=0.72$ and $\Omega_m=0.226$, both significantly lower
than in our WMAP7 cosmology. Indeed, their $\Omega_m$ value is even
lower than the one we assume for WMAP1. As a result, the differences
in halo mass function between their two cosmologies are much larger
than between our WMAP1 and WMAP7 models. {Recent work
by \cite{Kang2012} also finds very little variation of galaxy clustering
between the WMAP1 and WMAP7 cosmologies.}
  
In summary, with the WMAP7 cosmological parameters adopted here, we find
only small differences in galaxy properties relative to the WMAP1
model of G11, both in the local universe and at high redshift. This is
a consequence of the similar mass functions predicted by these two
cosmologies over the range of redshifts where galaxies form most of
their stars. The difference in cosmology is, in effect, too small to
show up strongly in either the evolution or the clustering of the
galaxies.  Given the substantial residual uncertainties in galaxy
formation modelling, it is not currently possible to distinguish the
two cosmologies using properties of the galaxy population. This may
reflect a degeneracy in the larger space of cosmological and galaxy
formation parameters. Our galaxy formation and simulation scaling
techniques make it feasible to combine low- and high-redshift
clustering and abundance data to constrain this larger parameter
space, and may eventually make it possible to seperate information about galaxy formation
processes from information about the larger cosmological context in
which they take place.

\section*{Acknowledgements}     

Galaxy, halo and lightcone catalogues corresponding closely to the
WMAP1 models of this paper are publicly available at
http://www.mpa-garching.mpg.de/millennium and similar catalogues for
the WMAP7 model will be made available on the same site as soon as
this paper is accepted for publication. REA, BH, GL and SW are
supported by Advanced Grant 246797 “GALFORMOD” from the European
Research Council. GQ acknowledges support from the National basic 
research program of China (program 973 under grant No. 2009CB24901), 
the Young Researcher Grant of National Astronomical Observatories, 
CAS, the NSFC grants program (No. 11143005) and the Partner Group 
program of the Max Planck Society. MB-K acknowledges support from the Southern California Center for Galaxy Evolution, a multi-campus research program funded by the University of California Office of Research. 

\bibliographystyle{mn2e}

\bibliography{draft}

\label{lastpage}
\end {document}